\documentclass[aps,showpacs,prb,twocolumn]{revtex4}
\usepackage{mathrsfs}
\usepackage{amssymb}
\usepackage{amsmath}
\usepackage{bm}
\usepackage{graphics}
\usepackage{natbib}

\newcommand{\sss}[1]{{\scriptscriptstyle #1}}
\begin{document}
\title{Coulomb-modified Fano interference in a double quantum dot Aharonov-Bohm ring}

\author{Weijiang Gong$^{a,b}$}
%\author{Yu Han$^{a,b}$}
\author{Xuefeng Xie$^{a}$}
\author{Guozhu Wei$^{a,b}$}\email[Corresponding author. Email address: ]
{guozhuwei02@sina.com}

\affiliation{ a. College of Sciences, Northeastern University,
Shenyang 110004, China \\
%b. Department of Physics, Liaoning University,
%Shenyang 110036, China \\
b. International Center for Material Physics, Acadmia Sinica,
Shenyang 110015, China}
\date{\today}

\begin{abstract}
%We carry out a comprehensive analysis of the Coulomb-modified Fano
%effect in electronic transport process through a double quantum dot
%Aharonov-Bohm ring.
In this paper, the Coulomb-induced changes of Fano interference in
electronic transport through a double quantum dot Aharonov-Bohm ring
are discussed. It is found that the Coulomb interaction in the
quantum dot in the reference channel can remarkably modify the Fano
interference, including the increase or decrease of the symmetry of
the Fano lineshape, as well as the inversion of the Fano lineshape,
which is dependent on the appropriate strength of the
Coulomb interaction. %But the nonzero Coulomb interaction
%only leads to the emergence of two-group Fano lineshapes.
When both the quantum dot levels are adjustable, the Coulomb-induced
splitting of the nonresonant channel leads to the destruction of the
Fano interference; whereas two blurry Fano lineshapes may appear in
the conductance spectra when the many-body effect in the dot of the
resonant channel is also considered. Interestingly, in the absence
of magnetic field, when the different-strength electron interactions
make one pair of levels of the dots in different channels the same,
the corresponding resonant state keeps vacuum despite the adjustment
of quantum dot levels.

\end{abstract}
% \keywords{}
 \maketitle

\bigskip

\section{Introduction}
The Fano effect, arising from the quantum interference between
resonant and nonresonant processes,\cite{refFano,Andrey} has been
observed in various physical fields, including neutron
scattering,\cite{Adair} atomic photoionization,\cite{Fano2} Raman
scattering,\cite{Cardona} optical absorption in quantum
wells,\cite{Faist} scanning tunneling microscopy,\cite{Chen} and
microwave scattering.\cite{Rotter} As a result, asymmetric
lineshapes appear in the spectra concerned, e.g., optical absorption
spectrum, usually called the Fano lineshapes.\cite{Andrey,Faist}
Quantum dots~(QDs), in particular, the coupled multiple QD
structures, provide multiple channels for electronic coherent
transmission. In appropriate parameter region, one or a few channels
serve as the resonant paths for electron tunneling and the others
are the nonresonant ones. Quantum interference of electron waves
going through these different paths inevitably leads to the
occurrence of Fano effect in the electron transport through these
structures.

\par
Experimentally, Fano resonances have been observed first in QDs at
the Kondo regime.\cite{Gores,Gordon} Subsequently, by embedding a
Coulomb blockaded QD in an Aharonov-Bohm (AB) ring interferometer, a
variety of Fano lineshapes were observed in the measured conductance
spectra. Conductance measurements exploring different geometries,
such as a quantum wire with a side-coupled
QD,\cite{refIye,refWang,refTor} a one-lead QD,\cite{Hanson} a ring
with side-coupled QD,\cite{Fuhrer} as well as the parallel double QD
structures,\cite{Blick,Chen2} provide more insight into the Fano
problem in mesoscopic systems.
\par
The occurrence of conductance dips in ballistic AB rings was
theoretically investigated almost 20 years ago.\cite{Weisz} Further,
early theoretical works examined the possibility of Fano lineshapes
in the transmission through one-dimensional waveguides and
waveguides with resonantly coupled cavities.\cite{Tekman} Recently,
inspired by the development of the relevant experimental works,
there have been a lot of theoretical investigations devoting
themselves to the Fano interference in electron transport through
various QD structures, for example, one or two QDs embedded in an AB
ring,\cite{refBulka,refKonig1,refKonig2,refGefen} double QDs in
different coupling
manners.\cite{refKang,refBai,refOrellana,refZhu,refDing} According
to these theoretical results, the Fano effect in QD structures
exhibits some peculiar behavoirs in electronic transport process, in
contrast to the conventional Fano effect. These include the tunable
Fano lineshape by the magnetic or electrostatic fields applied on
the QDs,\cite{refKang,refBai,refOrellana,refZhu} the Kondo resonance
associated Fano effect,\cite{refBulka,refKonig1,refDing,refMeden}
Coulomb-modification on the Fano effect,\cite{refZhu2} the
impurity-influenced Fano interference,\cite{gongjap} the relation
between the dephasing time and the Fano parameter
$q$,\cite{refClerk} and the spin-dependent Fano interference when
various spin-relevant field is applied.\cite{Chi}
\par
According to these previous works, since the understanding of the
important role of Coulomb interactions in electron transport through
the coupled-QD structures, in so many previous
literatures,\cite{refBulka,refKonig1,refDing,refMeden} when the Fano
interference in electron tunneling through the corresponding QD
structures were investigated, the influence of the many-body effect
on the Fano resonance is always a leading concern. Albeit these
theoretical descriptions, some aspects about the Coulomb-modified
Fano effect in QD structures deserve further theoretical
investigation. First, we have to know that in the mixed-valence
regime, the Coulomb interactions also contribute nontrivially to the
electron transport process, since the quantum interference is
modified by the Coulomb-induced splitting of the QD levels
(Interpretatively, $\varepsilon_0$ is changed into $\varepsilon_0$
and $\varepsilon_0+U$ with $\varepsilon_0$ being one QD level and
$U$ the Coulomb interaction strength). Accordingly, the
Coulomb-induced multi-channel quantum interference presents some
complicated properties quite different from those in the
noninteracting case. Such an issue was discussed in the model of
parallel-coupled double QDs, in which the observable change of the
Fano effect is exhibited.\cite{refZhu2} Secondly, as is known, with
respect to the multi-QD structures, such as the double-QD AB ring,
it is not necessary that there are equal electron interactions in
the respective QDs.\cite{Silva} Thereby, the investigation of the
influence of the unequal Coulomb repulsions on the quantum
interference, e.g., the Fano interference, is desirable.
\par
Motivated by such a topic, in this work we concentrate our attention
on the Coulomb-modified Fano effect in electronic transport through
a double-QD AB ring. Then, with the help of the standard Fano form
of the linear conductance expression, the Coulomb-induced changes of
the Fano lineshapes in the linear conductance spectra are discussed
in detail. It is found that the Coulomb interaction in the QD in the
reference channel plays a nontrivial role in the change of the Fano
lineshapes. For the case of the adjustable levels of both QDs, the
Coulomb-induced splitting of the nonresonant channel leads to the
destruction of the Fano interference. Only when the many-body effect
in both QDs of the ring, blurry Fano lineshapes are possible to
appear in the conductance spectra. In addition, in the
zero-magnetic-field case, when the different-strength electron
interactions make any pair of QD levels of different channels
consistent, the corresponding resonant state keeps vacuum.
\par
The rest of the paper is organized as follows. In Sec. \ref{theory},
the model Hamiltonian to describe the electron motion in double-QD
AB ring is introduced firstly. Then a formula for the linear
conductance is derived by means of the nonequilibrium Green function
technique and the Fano form of the conductance expression are
obtained. In Sec. \ref{result2}, the calculated results about the
linear conductance spectrum are shown. Then a discussion focusing on
the change of Fano lineshape is given. Finally, the main results are
summarized in Sec. \ref{summary}.

\section{model\label{theory}}
The double-QD AB ring we consider is illustrated in Fig.
\ref{structure}(a). The Hamiltonian to describe the electronic
motion in this structure reads $H=H_{C}+H_{D}+H_{T}$. The first term
is the Hamiltonian for the noninteracting electrons in the two
leads:
\begin{equation}
H_{C}=\underset{\sigma,k,\alpha\in L,R}{\sum }\varepsilon _{\alpha
k}c_{\alpha k\sigma}^\dag c_{\alpha k\sigma },\label{2}
\end{equation}
where $c_{\alpha k\sigma}^\dag$ $( c_{\alpha k\sigma})$ with
$\sigma=\pm1$ (or $\uparrow, \downarrow$) being the spin index is an
operator to create (annihilate) an electron of the continuous state
$|k,\sigma\rangle$ in lead-$\alpha$, and $\varepsilon _{\alpha k}$
is the corresponding single-particle energy. The second term
describes electron in the two QDs in the arms of the ring, which
takes a form as
\begin{equation}
H_{D}=\sum_{j=1,\sigma}^{2}\varepsilon _{j}d_{j\sigma}^\dag
d_{j\sigma}+\sum_{j=1}^{2}U_jn_{j\uparrow}n_{j\downarrow},\label{3}
\end{equation}
where $d^{\dag}_{j\sigma}$ $(d_{j\sigma})$ is the creation
(annihilation) operator of electron in QD-$j$. And $\varepsilon_j$
denotes the electron level in the corresponding QD, while $U_j$
represents the intradot Coulomb repulsion. We assume that only one
level is relevant in each QD. The last term in the Hamiltonian
describes the electron tunneling between the leads and QDs. It is
given by
\begin{equation}
H_{T} =\underset{jk,\sigma}{\sum }( V_{j\sss{L}}d_{j\sigma}^\dag
c_{\sss{L}k\sigma}+ V_{j\sss{R}}d_{j\sigma}^\dag
c_{\sss{R}k\sigma}+\mathrm {H.c.}), \label{4}
\end{equation}
where $V_{j\alpha}$ denotes the QD-lead coupling strength. The
tunnelling matrix elements take the following values:
$V_{1\sss{L}}=|V_{1\sss{L}}|e^{i\phi/4}$,
$V_{1\sss{R}}^*=|V_{1\sss{R}}|e^{i\phi/4}$,
$V_{2\sss{R}}=|V_{2\sss{R}}|e^{i\phi/4}$, and
$V_{2\sss{L}}^*=|V_{2\sss{L}}|e^{i\phi/4}$. The phase shift $\phi$
is associated with the magnetic flux $\Phi$ threading the system by
a relation $\phi=2\pi\Phi/\Phi_{0}$, in which $\Phi_{0}=h/e$ is the
flux quantum.
\par
To study the electronic transport properties in the linear regime,
the linear conductance at zero temperature is obtained by the
Landauer-B\"{u}ttiker formula\cite{refLandauer,refMeir1}
\begin{equation}
\mathcal {G}=\frac{e^{2}}{h}\sum_\sigma
T_\sigma(\omega)|_{\omega=\varepsilon_F}.\label{conductance}
\end{equation}
$T_{\sigma}(\omega)$ is the transmission function, in terms of Green
function which takes the form as $T_{\sigma}(\omega)=\mathrm
{Tr}[\Gamma^LG^r_{\sigma}(\omega)\Gamma^RG^a_{\sigma}(\omega)]$,
where $\Gamma^L$ is a $2\times 2$ matrix and defined as
$[\Gamma^{L}]_{jn}=2\pi
V_{j\sss{L}}V^*_{n\sss{L}}\rho_\sss{L}(\omega)$, describing the
coupling strength between the two QDs and lead-$L$. We will ignore
the $\omega$-dependence of $\Gamma^{L}_{jn}$ since the electron
density of states in lead-$L$, $\rho_\sss{L}(\omega)$, can be
usually viewed as a constant. By the same token, we can define
$[\Gamma^R]_{jn}$. Besides, the retarded and advanced Green
functions in Fourier space are involved here. By means of the
equation-of-motion method and via a straightforward
derivation,\cite{refGong} we obtain the retarded Green functions
written in a matrix form:
\begin{eqnarray}
G^r_{\sigma}(\omega)=\left[\begin{array}{cc} g_{1\sigma}(z)^{-1} & i\Gamma_{12}\\
  i\Gamma_{21}& g_{2\sigma}(z)^{-1}
\end{array}\right]^{-1}\ \label{green},
\end{eqnarray}
with $z=\omega+i0^+$. And
$g_{j\sigma}(z)=[(z-\varepsilon_{j})\Lambda_{j\sigma}+i\Gamma_{jj}]^{-1}$,
is the zero-order Green function of the QD-$j$ unperturbed by
another QD, where $\Gamma_{jn}={1\over
2}\{[\Gamma^L]_{jn}+[\Gamma^R]_{jn}\}$. Besides,
\begin{equation}
\Lambda_{j\sigma}=\frac{z-\varepsilon_{j}-U_j}{z-\varepsilon_{j}-U_j+U_j\langle
n_{j\bar{\sigma}}\rangle}
\end{equation}
results from the second-order approximation (i.e., Hubbard
approximation) for Coulomb terms.\cite{refGong,refLiu} The average
electron occupation number in QD-$j$ is determined by the relation
$\langle n_{j\sigma}\rangle=\frac{i}{2\pi}\int d\omega
[{G}^{<}_\sigma(\omega)]_{jj}$, in which
${G}^{<}_\sigma(\omega)={G}^{r}_\sigma(\omega)\Sigma^<{G}^{a}_\sigma(\omega)$
with $\Sigma^<=i\Gamma^Lf_L+i\Gamma^R f_R$. In addition, the
advanced Green function can be readily obtained via a relation
$G_{\sigma}^a(\omega)=[G_\sigma^r(\omega)]^\dag$.
\par
With the solution of the Green function and the definition of the
coupling matrixes $\Gamma^\alpha$, we can express the linear
conductance defined by Eq.(\ref{conductance}) in terms of the
structure parameters, i.e.,
\begin{widetext}
\begin{equation}
\mathcal
G=\frac{e^2}{h}\sum_{\sigma}\frac{\tilde{\varepsilon}^2_{2\sigma}\Gamma^L_{11}\Gamma^R_{11}+\tilde{\varepsilon}^2_{1\sigma}\Gamma^L_{22}\Gamma^R_{22}+\tilde{\varepsilon}_{1\sigma}\tilde{\varepsilon}_{2\sigma}
(\Gamma^L_{12}\Gamma^R_{21}+\Gamma^L_{21}\Gamma^R_{12})}
{(\tilde{\varepsilon}_{1\sigma}\tilde{\varepsilon}_{2\sigma}-\Gamma_{11}\Gamma_{22}+\Gamma_{12}\Gamma_{21})^2
+(\Gamma_{11}\tilde{\varepsilon}_{2\sigma}+\Gamma_{22}\tilde{\varepsilon}_{1\sigma})^2},
\label{expression}
\end{equation}
\end{widetext}
with
$\tilde{\varepsilon}_{j\sigma}=\varepsilon_{j}\Lambda_{j\sigma}$
being the renormalized level of QD-$j$. The expression corresponds
to Eq. (\ref{expression}) in the work of C. Karrasch $et$
$al.$\cite{refMeden} distinctly. In order to study the Fano
interference, one usually rewrites the conductance expression into a
Fano form, i.e.,
\begin{equation} \mathcal
G=\frac{e^2}{h}\sum_\sigma
T_{b\sigma}\frac{|e_\sigma+q_\sigma|^2}{e^2_\sigma+1},
\label{fanoform}
\end{equation}
in which the three auxiliary quantities are defined as
$T_{b\sigma}=\Gamma^L_{11}\Gamma^R_{11}/[|\tilde{\varepsilon}_{1\sigma}|^2+\Gamma^2_{11}]$,
$e_\sigma=-\mathrm {Re}G^r_{22,\sigma}/\mathrm {Im}G^r_{22,\sigma}$,
and $q_\sigma=-\frac{\tilde{\varepsilon}_{1\sigma}}{\Gamma_{11}}
(\Gamma^L_{12}\Gamma^R_{21}\Gamma_{11}-
T_{b\sigma}\Gamma_{12}\Gamma_{21}\Gamma_{11})/({\Gamma^L_{11}
\Gamma^R_{11}\Gamma_{22}-T_{b\sigma}\Gamma_{12}\Gamma_{21}\Gamma_{11}})$.
Obviously, $T_{b\sigma}$ is the ability of electron transmission
through the upper arm of the ring. Under the condition of symmetric
QD-lead coupling, $\Gamma^L_{jj}=\Gamma^R_{jj}$, so $q_\sigma$ can
be simplified as
$q_\sigma=-\frac{\tilde{\varepsilon}_{1\sigma}}{\Gamma_{11}}(e^{i\phi}-T_{b\sigma}
\cos^2\frac{\phi}{2})/(1-T_{b\sigma} \cos^2\frac{\phi}{2})$. Also,
from Eq.(\ref{fanoform}) we can find that Fano antiresonance emerges
while
$e_{\sigma}+q_{\sigma}=-[\Gamma^L_{11}\Gamma^R_{11}\tilde{\varepsilon}_{2\sigma}
+\Gamma^L_{12}\Gamma^R_{21}\tilde{\varepsilon}_{1\sigma}]/
[4\Gamma^L_{11}\Gamma^R_{11}\Gamma_{22}-T_{b\sigma}\Gamma_{11}\Gamma_{12}\Gamma_{21}]=0$,
which indicates that Fano antiresonance emerges inevitably while
$\Gamma^L_{11}\Gamma^R_{11}\tilde{\varepsilon}_{2\sigma}+\Gamma^L_{12}\Gamma^R_{21}\tilde{\varepsilon}_{1\sigma}=0$.

\section{Numerical results and discussions \label{result2}}
Based on the formulation developed in the previous section, we can
then carry out the numerical calculation to investigate the Fano
interference in electron transport through such a double-QD AB ring
structure. Before proceeding, we consider $\Gamma$ as the unit of
energy and take the Fermi level of the system $\varepsilon_F$ as the
zero point of energy.
\par
First, with the help of a standard Fano expression
Eq.({\ref{fanoform}), by analyzing the Fano parameter $q_\sigma$ we
can investigate the appearance of the Fano lineshape in the linear
conductance spectrum. It is certain that when $q_\sigma$ is real,
the linear conductance is possible to display a standard Fano
lineshape, and moreover the change of the sign of $q_\sigma$ ($+/-$)
can lead to the reversal of the Fano lineshape. As discussed in the
previous works,\cite{refOrellana,refZhu,gongpe} there are two ways
to realize the change of the sign of $q_{\sigma}$, i.e., by choosing
the different-sign values of the level of QD-1 with respect to the
Fermi level or tuning the threading magnetic flux from $\phi=2n\pi$
to $\phi=(2n+1)\pi$ ($n\in Integer$). As shown in Fig.1(b) and (c),
when the level of QD-1 is fixed at a nonzero value with respect to
the Fermi level, the upper arm of the AB ring accordingly provides a
reference (nonresonant) channel for the Fano interference,
consequently, with the shift of the level of QD-2 the calculated
conductance spectra present the Fano lineshapes apparently. To be
concrete, in the absence of magnetic flux, $q_{\sigma}$ is equal to
$-{\varepsilon_1\over \Gamma_{11}}$, thereby, the opposite-sign
values of QD-1 level causes the opposition of signs ($+/-$) of
$q_{\sigma}$, which gives rise to the inversion of the Fano
lineshape. Of course, $\varepsilon_1=0$, implying $q_{\sigma}=0$, is
a critical position of the lineshape's inversion, correspondingly,
in such a case the liner conductance is equal to $2e^2/h$,
independent of the modulation of $\varepsilon_2$. On the other hand,
when the magnetic flux increases to $\pi$, the Fano parameter
$q_{\sigma}$ has the expression $q_{\sigma}={\varepsilon_1\over
\Gamma_{11}}$, hence the Fano lineshape can also be inverted despite
$\varepsilon_1=-{\Gamma\over 2}$ fixed, as shown in
Fig.\ref{structure}(c).

\par
Next, we turn to pay attention to the influence of the many-body
effect on the change of the Fano lineshape. It is known that the
many-body effect is an important origin for the peculiar transport
properties in coupled-QD structures. Therefore, it is supposed to
influence the Fano resonance to some extent. Usually, the many-body
effect is incorporated by considering only the intradot Coulomb
repulsion, i.e., the Hubbard term. In general, if the Hubbard
interaction is not very strong, we can truncate the equations of
motion of the Green functions to the second
order.\cite{refGong,refLiu} In Figs.\ref{highorder}, by choosing
$U_2=0$ and $\varepsilon_2=\varepsilon_0+{\Gamma\over 2}$, we first
investigate the influence of the many-body effect in QD-1 on the
Fano interference and plot the Coulomb-modified linear conductance
spectra as functions of $\varepsilon_2$. From the figure, we can
readily see that the increase of Coulomb repulsion strength of QD-1
indeed changes the Fano lineshape in a nontrivial way. As a typical
case, in Fig.\ref{highorder}(a) with $\varepsilon_1=-{\Gamma\over
2}$, when $U_1=\Gamma$ it is interesting that the conductance
spectrum presents a Breit-Wigner lineshape with its resonant peak at
the point of $\varepsilon_2=0$, corresponding to the dotted line in
this figure. This indicates that in such a case, only the electron
transport through the down arm of the ring is allowed, so that the
Fano interference vanishes completely. With the increment of the
Coulomb strength, such as for the cases of $U_1=2\Gamma$ and
$3\Gamma$, the conductance profiles show themselves as the Fano
lineshapes again, but the asymmetry degree of them is weaker
compared with that in the noninteracting case. Next, in the cases of
$\varepsilon_1=-\Gamma$, as exhibited in Fig.\ref{highorder}(b),
when $U_1=\Gamma$ the corresponding conductance is equal to
$2e^2\over h$ in the whole range of $\varepsilon_2$; but when
$U_1=2\Gamma$, the linear conductance spectrum gets close to the
Breit-Wigner lineshape as well, similar to the result in the case of
$\varepsilon_1=-{\Gamma\over 2}$ and $U_1=\Gamma$. Alternatively,
under the condition of $\varepsilon_1=-2\Gamma$ it is shown that the
increase of $U_1$ (i.e., the case of $U_1=3\Gamma$) can give rise to
the reversal of the Fano lineshape, just as displayed by the
shot-dashed line in Fig.\ref{highorder}(c). Therefore, by virtue of
the above results, the effect of the Coulomb interaction in the
reference channel on the Fano interference is striking.
\par
We can find, from Eq.(\ref{expression}), that in the linear
transport regime, the many-body effect modify the electron transport
through such a structure by renornalizing the QD levels
$\varepsilon_j$ to
$\tilde{\varepsilon}_{j\sigma}=\varepsilon_j\Lambda_{j\sigma}=\varepsilon_j(i0^+-\varepsilon_{j}-U_j)/(i0^+-\varepsilon_{j}-U_j+U_j\langle
n_{j\bar{\sigma}}\rangle)$. Therefore, it is understood that the
many-body effect in QD-1 can efficiently adjust the Fano lineshape
in the conductance curves with its modulating the Fano parameter
$q_{\sigma}$, since in the absence of magnetic flux
$q_{\sigma}=-{\varepsilon_{1\sigma}\over
\Gamma_{11}}=-{\varepsilon_1\over \Gamma_{11}}\Lambda_{1\sigma}$ (
or $q_{\sigma}={\varepsilon_{1\sigma}\over
\Gamma_{11}}={\varepsilon_1\over \Gamma_{11}}\Lambda_{1\sigma}$ when
$\phi=\pi$). In Fig.\ref{fanoline}(a) and (b), by assuming
$\varepsilon_1=-{\Gamma\over 2}$, $-\Gamma$, and $-2\Gamma$, the
profiles of ${\rm Re}[\Lambda_{1\sigma}]$ and ${\rm
Im}[\Lambda_{1\sigma}]$ are plotted as functions of $U_1$,
respectively. It is seen that in general regime, only the real part
of $\Lambda_{1\sigma}$ contributes to the variation of $q_{\sigma}$,
whereas when the condition of $U_1=2\varepsilon_1$ is satisfied,
both the real and imaginary parts of $\Lambda_{1\sigma}$ play
important roles in the sharp change of $q_{\sigma}$. Taking the case
of $\varepsilon_1=-{\Gamma\over 2}$ as an example, in the vicinity
of $U_1=\Gamma$, $|\Lambda_{1\sigma}|$ approximately goes to
infinity, so, in such a case $q_{\sigma}\rightarrow\infty$ and then
the conductance expression can mathematically be simplified as
${\cal G}={e^2\over h}\sum_{\sigma}{1\over e^2_{\sigma}+1}={e^2\over
h}\sum_{\sigma}|g_{2\sigma}|^2$, which is according to the display
of a Breit-Wigner lineashape in the linear conductance curve. The
physical reason of such a result can be clarified as follows. As is
known, in the presence of many-body effect, the level of QD-1 splits
into two, i.e., $\varepsilon_1$ and $\varepsilon_1+U_1$, thus the
consideration of $\varepsilon_1=-{\Gamma\over 2}$ and $U_1=\Gamma$
gives rise to the electron-hole symmetry of QD-1 and the occupation
number of $\sigma$-spin in QD-1 is just $\langle n_{1\sigma}\rangle
={1\over 2}$. By virtue of the description in the previous
works,\cite{refGong} we can be sure that the electron-hole symmetry
is able to restrain the electron tunneling through the upper arm of
the ring, when the electron correlation effect is ignored. As a
consequence, in such a case only the electron transmission through
the down arm ( resonant channel ) is allowed, which results in the
appearance of the Breit-Wigner lineshape in the linear conductance.
On the other hand, from Fig.\ref{fanoline}(a) and (b), it can also
be found that when $\varepsilon_1=-\Gamma$ and $U_1=\Gamma$ is
considered, $\Lambda_{1\sigma}$ becomes equal to zero, hence in such
a case the Fano parameter $q_{\sigma}=0$ and $e_{\sigma}\rightarrow
\infty$, as a result, the conductance is equal to $2e^2\over h$
independent of the tuning of $\varepsilon_2$. On the other hand, for
the case of $\varepsilon_1=-2\Gamma$ and $U_1=3\Gamma$, one can see
that $\Lambda_{1\sigma}$ becomes negative obviously, which results
in the inversion of the Fano lineshape in the calculated conductance
profile.
\par
Typically, let us continue to devote ourselves to the curve of ${\rm
Re}[\Lambda_{1\sigma}]$ and ${\rm Im}[\Lambda_{1\sigma}]$ of
$\varepsilon_1=-{\Gamma\over2}$. One can readily find that in the
process of $U_1$ increasing to $\Gamma\over 2$, ${\rm
Re}[\Lambda_{1\sigma}]$ decreases from $1$ to zero, which leads to
the decrease of $q_{\sigma}$ ( since here ${\rm
Im}[\Lambda_{1\sigma}]=0$ ) and the increase of the asymmetry of the
Fano lineshape, corresponding to the results in
Fig.\ref{fanoline}(c). Well, it is clear that $U_1={\Gamma\over 2}$
make $q_{\sigma}$ close to infinity, so the conductance becomes a
constant irrelevant to the shift of $\varepsilon_2$. When the
Coulomb repulsion strength exceeds $\Gamma\over 2$ $\rm
Re[\Lambda_{1\sigma}]$ becomes negative until the point of
$U_1=\Gamma$, accordingly the sign of $q_{\sigma}$ changes and such
a result brings about the reversal of the Fano lineshape in the
calculated linear conductance spectrum, as shown in
Fig.\ref{fanoline}(d). When go on increasing the Coulomb strength in
QD-1, we can see that $\rm Re{\Lambda_{1\sigma}}$ always keeps
positive with the amplitude of it greater than one, so under such a
condition the weak modulation of $U_1$ on the Fano lineshape is well
understood, too. With the results above, we can clarify the
phenomenon of the modulation of many-body effect in QD-1 on the Fano
interference in such a system. In addition, it is necessary to point
out that when the Coulomb interaction in QD-2 is taken into account,
the electron interaction in QD-2 can induce the splitting of the
level of QD-2(i.e., $\varepsilon_2$ and $\varepsilon_2+U_2$), so it
is clear that the Fano lineshape in the conductance spectrum will be
divided into two groups, but the properties of the Fano lineshape
are still determined by the Coulomb term in QD-1, as shown in
Fig.\ref{fanoline}(e).
\par
For the case of the adjustable $\varepsilon_1$, by letting
$\Gamma_{11}=10\Gamma_{22}=\Gamma$ and
$\varepsilon_1=\varepsilon_0-{\Gamma\over2}$ one can also readily
find a Fano lineshape in the linear conductance spectrum, as shown
in the inset of Fig.\ref{fanoline1}(a). This is for the reason that
here the upper arm of this ring can provide a `less' resonant path
while its down arm provides a `more' resonant path for the
occurrence of Fano interference. Then, according to the results in
Fig.\ref{fanoline1} (a) and (b), in comparison with that in the
noninteracting case, in such a case the existence of the many-body
effect in QD-1 can also modify the Fano interference to a great
extent. In the absence of magnetic flux and the many-body term in
QD-2, as shown in Fig.\ref{fanoline1}(a), we find that when
$U_1=\Gamma$, in the conductance spectrum there exist two peaks
respectively corresponding to the positions of
$\varepsilon_0={\Gamma\over 2}$ and $-{\Gamma\over 2}$ with the
approximately-equal width of them, and at the vicinity of
$\varepsilon_0=0$ the conductance encounters its zero value. It is
therefore certain that in such a case, the original Fano
interference disappears completely since the consideration of
many-body effect. Such a result can be well explained as follows.
Certainly, the Coulomb interaction in the Hubbard approximation can
lead to the splitting of the energy level $\varepsilon_1$ into
$\varepsilon_1$ and $\varepsilon_1+U_1$,\cite{refGong,refLiu} so
with the adjustment of $\varepsilon_1$ in the conductance curve
there are two peaks corresponding to the points of $\varepsilon_1=0$
and $\varepsilon_1+U_1=0$, respectively. With respect to the zero
point of the conductance, we can understand that it arises from the
electron-hole symmetry, as discussed above. But, we have to note
that in this case due to $\varepsilon_2=\varepsilon_1+U_1$, the
conductance peak at the point of $\varepsilon_1+U_1=0$ originates
from the constructive interference between the electron waves
passing through the two arms, since at this position the absence of
magnetic flux causes the uniform phase of the electron waves. With
the increase of $U_1$, there appear three conductance peaks in the
linear conductance spectra, respectively associated with the points
of $\varepsilon_1$, $\varepsilon_2$, and $\varepsilon_1+U_1$ equal
to the Fermi level. Apparently, we have to know that owing to the
Coulomb-induced splitting of the conductance spectrum in the upper
arm, electron transmission though this channel can not be regarded
as a `less' resonant process. Therefore, the Fano interference is
seriously destroyed and the linear conductance spectrum does not
display a standard Fano lineshape any more.
\par
When a local magnetic flux is introduced with $\phi=\pi$, as shown
in Fig.\ref{fanoline1}(b), it is found that for the case of
$U_1=\Gamma$, opposite to the result of zero magnetic field, when
$\varepsilon_2$ (i.e., $\varepsilon_1+U_1$) coincides with the Fermi
level of the system the electron transport becomes antiresonant.
This is because that the application of magnetic field with
$\phi=\pi$ gives rise to the opposition of the phases of the
electron waves traveling through the two arms, and the destructive
interference results in the occurrence of antiresonance. But notice
that this antiresonance can not be viewed as the Fano antiresonance.
Then, even if increasing the Coulomb interaction strength in QD-1,
one can find that there is still no Fano lineshape in the
conductance spectrum, especially for the case of $U_1=2\Gamma$,
distinctly no Fano antiresonance valley exists in the conductance
spectrum, which can also be obtained analytically.

\par
In Fig. \ref{fanoline1}(c) and (d) the many-body effect in QD-2 is
also taken into account, and it shows the conductance spectra of
$U_2=\Gamma$. Here, since the nonzero $U_2$, the level of QD-2
$\varepsilon_2$ splits into two, i.e., $\varepsilon_2$ and
$\varepsilon_2+U_2$, the electron transport through the down arm can
result in two resonant peaks at the positions of $\varepsilon_2=0$
and $\varepsilon_2+U_2=0$, respectively, as shown by the dashed line
in Fig.\ref{fanoline1}(c) with $U_1=\Gamma$. Clearly, the
conductance peak corresponding to the point of
$\varepsilon_0=-{\Gamma\over2}$ originates from the constructive
interference of the electron waves in the two arms; when the
Coulomb-induced level $\varepsilon_2+U_2$ is consistent with the
Fermi level, the electron traveling through the down arm of this
ring can provide a `more' resonant path while the upper arm provides
a `less' resonant path, so in this regime the linear conductance
spectrum also presents a Fano lineshape. Next, when the Coulomb
interaction strength is increased to $U_1=2\Gamma$, the level
$\varepsilon_2+U_2$ has an opportunity to coincide with the level
$\varepsilon_1+U_1$, thus when $\varepsilon_0$ is adjusted to the
position of $\varepsilon_0=-{3\Gamma\over 2}$, the coherent electron
transport shows a resonant peak in the case of $\phi=0$, whereas the
conductance becomes zero when the magnetic flux phase factor
increases to $\phi=\pi$. On the other hand, when the Coulomb
repulsion in QD-1 is taken to be $U_1=3\Gamma$, the conductance
profile is divided into two groups symmetric about the axis of
$\varepsilon_0=-\Gamma$, where the electron-hole symmetry just comes
into being. And in each group there is a Fano lineshape,
respectively around the positions of $\varepsilon_2=0$ and
$\varepsilon_2+U_2=0$. It is noteworthy that, when the magnetic
phase factor $\phi=\pi$, an insulating band comes up obviously
between the two Fano peaks. This is for the reason that in this
region the conductance zero induced by the electron-hole symmetry
and two Fano antiresonance occurs, so the conductance in this region
is seriously suppressed. And a similar discussion can be found in
Ref[\onlinecite{refGong}].

\par
In the following, let us turn to focus on the average electron
occupation numbers in the two QDs, as displayed in
Fig.\ref{orders}(a)-(d). First, as shown in Fig.\ref{orders}(a),
remarkably different from the results in the other cases, it is
interesting that with regard to the case of $\phi=0$, when only the
electron interaction in QD-1 is considered $U_1=\Gamma$, the average
electron occupation number in QD-2 is seriously limited (i.e.,
$\langle n_{2\sigma}\rangle\rightarrow 0$), independent of the
adjustment of gate voltage around the Fermi level. So in such a
structure, by the presence of appropriate conditions the zero
electron occupation ( i.e., the vacuum state ) in QD-2 can be
achieved in principle. In order to uncover such a result, there is
need to analyze the local density of states (LDOS) in QD-2. Just as
shown in Fig.\ref{compare}(a), we investigate the LDOS spectrum of
QD-2 with the relevant quantities fixed at $\varepsilon_0=-\Gamma$,
$U_1=\Gamma$, and $U_2=0$. It is found that in the case of zero
magnetic flux, the LDOS in QD-2 keeps zero in the whole range of
$\omega$, accordingly one can understand the result of zero electron
occupation number in QD-2. Such a result can be explained
analytically by paying attention to the Green function
$G^r_{22,\sigma}$, since the LDOS in QD-2 can be written out
following the formula $\rho_{2\sigma}=-{1\over
\pi}\rm{Im}G^r_{22,\sigma}$. We then first obtain the analytical
expression of $G^r_{22,\sigma}$ with
$G^r_{22,\sigma}={g_{2\sigma}\over
1+g_{1\sigma}\Gamma_{12}g_{2\sigma}\Gamma_{21}}$. Due to the level
of QD-2 $\varepsilon_2$ identical with $\varepsilon_1+U_1$, the
expression $\rho_{2\sigma}$ can be expressed explicitly as
$\rho_{2\sigma}={1\over \pi}{\Gamma_{22}\over
(\omega-\varepsilon_2)^2+(\frac{\Gamma_{11}}{2}+\Gamma_{22})^2}$
when the magnetic flux is absent (Here the strength of the coupling
between the level $\varepsilon_1+U_1$ and the leads are assumed to
be $\Gamma_{11}\over2$ for simplicity). Because of
$\Gamma_{11}\gg\Gamma_{22}$, one can well understand the result of
$\rho_{2\sigma}\rightarrow 0$. To be contrary, we can see that when
$\phi=\pi$, $\rho_{2\sigma}={1\over \pi}{\Gamma_{22}\over
(\omega-\varepsilon_2)^2+\Gamma_{22}^2}$, which leads to the clear
peak of the LDOS spectrum, as shown by dotted line in
Fig.\ref{compare}(a). The further explanation about this result
should fall back on the analysis of the quantum interference in this
system by means of the language of Feynman path.\cite{gongpe} Next,
for the reason alike, we can understand that at the zero magnetic
field case, when $U_1=U_2=\Gamma$ $\langle n_{2\sigma}\rangle$
begins to increase sharply at the point of
$\varepsilon_0=-{3\Gamma\over 2}$, since the identical
$\varepsilon_2$ and $\varepsilon_1+U_1$, as shown in
Fig.\ref{orders}(c). We might as well conclude that in such a
structure, in the absence of magnetic field, when a localized state
is completely wrapped by a expanding state, the average electron
occupation number in it will be close to zero and such a state will
become vacuum. Thus, under the condition of $U_1=2\Gamma$ and
$U_2=\Gamma$, the average occupation number of $\sigma$-spin
electron in QD-2 has its maximum $1\over 2$ [see Fig.\ref{orders}(c)
and Fig.\ref{compare}(b)], since in such a case the Coulomb-induced
levels in the two QDs are the same as each other.
\par
Lastly, we have to mention the influence of the electron
interactions on the Fano interference when the electron correlation
is taken into account. Well, when the electron interaction is very
strong, one need extend the theoretical treatment by adding the
interdot interaction and beyond the second order approximation. Then
the further modification to the Fano lineshape will naturally arise.
For example, when the QD in the upper arm is in the Kondo regime,
there will be a bound state aligned with the Fermi level of the
system.\cite{Kondo,Kondo2} Thereby, the Kondo resonance will occur
in the upper arm ( the reference channel ) of such a structure and
accordingly the Fano parameter will take a value close to zero.
Thus, we can predict that in the Kondo regime, the Fano lineshape in
the linear conductance spectrum will be modulated to a great degree.
Surely, the renormalized group (RG)
technique,\cite{refWillson,refSindel,refMeden2} is an appropriate
method to deal with this problem. SO, with this idea we can
investigate this interesting subject in the future.

\par

\par

\section{summary\label{summary}}
To sum up, in this work we have systematically studied the
Coulomb-modified Fano effect in electron transport process through a
double-QD AB ring. Firstly, we established an expression of the
linear conductance in the standard Fano form. Then, the change of
the Fano parameter $q_{\sigma}$ has received much attention by the
presence of the many-body effect in the QDs, and based on the
obtained results, the changes of the Fano lineshapes in the linear
conductance spectra have been well investigated. Consequently, we
have found that the Coulomb interaction in QD-1 (i.e., the dot in
the reference channel) contributes much to the change of the Fano
interference, and furthermore appropriate Coulomb interaction can
lead to the inversion of the Fano lineshape. But the nonzero Coulomb
interaction in QD-2 (the dot of the resonant channel) only brings
about the emergence of two-group Fano lineshapes, the variation of
which are still determined by the electron interaction in QD-1.
\par
When both the QD levels are adjustable with respect to the Fermi
level, the Coulomb-induced splitting of the nonresonant channel
gives rise to the destruction of the Fano interference. Only for the
cases of the many-body effect in the QD of the resonant channel also
being considered, two blurry Fano lineshapes emerge in the
conductance spectra again. In addition, we have also found that the
difference of the strength of the electron interactions in the two
QDs has a opportunity to make one pair of QD levels of the two
channels consistent with each other, which causes the corresponding
resonant state vacuum in the absence of magnetic field. And by
analyzing the LDOS in the QD in the resonant channel, such a result
is clarified. We hope that all these results could be helpful for
the relevant experiment.

\section*{Acknowledgments}
This work was financially supported by the National Natural Science
Foundation of China (Grant No. 10847109). %and the Scientific Research
%Project of Liaoning Education Office (Grant No. 2009A309).
%\appendix* \section{a proof}

\clearpage
%\section{\protect\bigskip\ {\protect\large FIGURES}}

\bigskip
\begin{figure}
\centering \scalebox{0.35}{\includegraphics{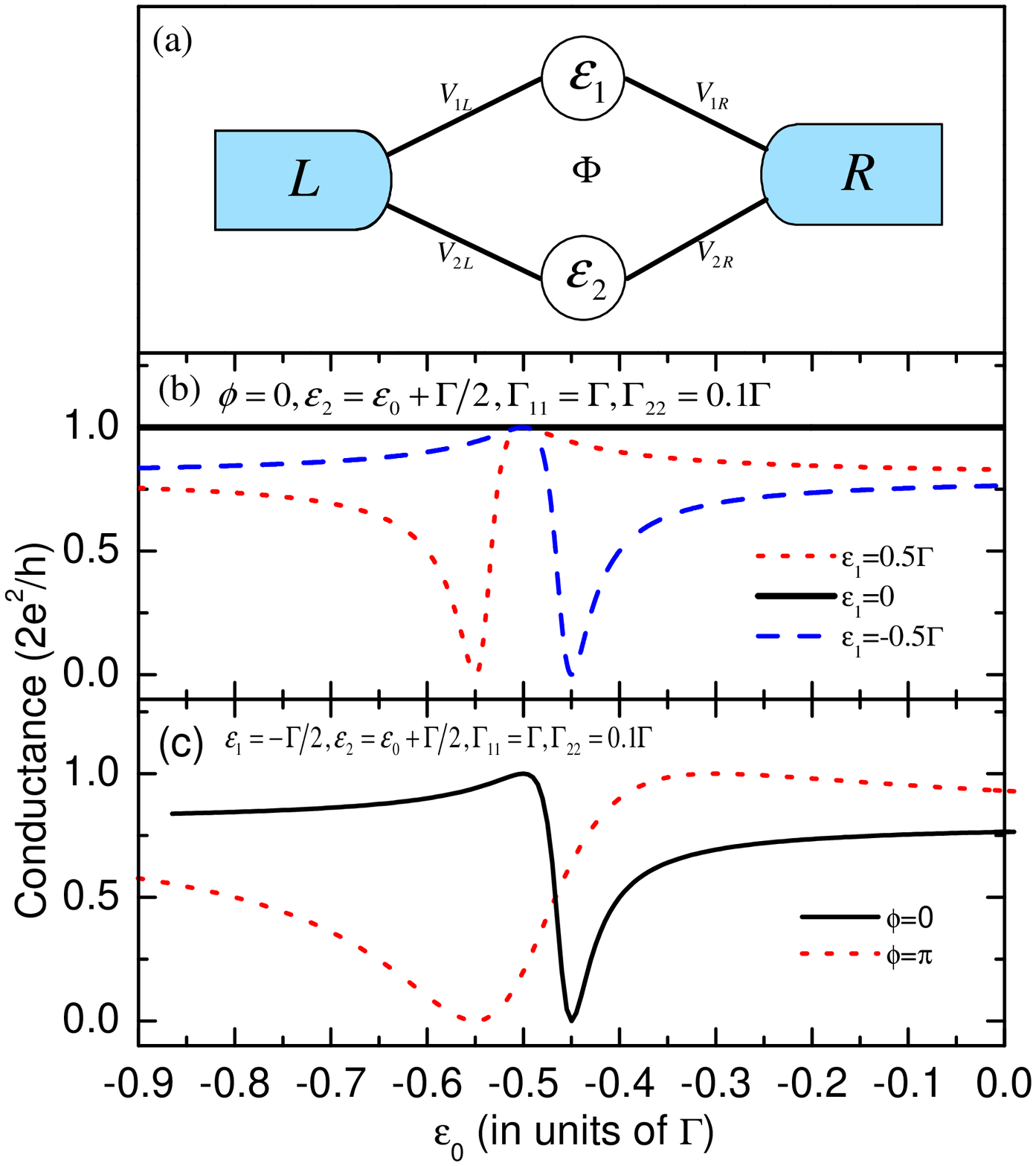}} \caption{
(a) Schematic of a double-QD AB interferometer. (b) The linear
conductance spectra with $\varepsilon_1=\frac{1}{2}\Gamma$, $0$, and
$-\frac{1}{2}\Gamma$, respectively, in the absence of magnetic
field. (c) The compared conductance spectra for the cases of
$\phi=0$ and $\pi$ with $\varepsilon_1=-\frac{1}{2}\Gamma$. The
QD-lead couplings are $\Gamma_{11}=10\Gamma_{22}=\Gamma$ and and the
level of QD-2 $\varepsilon_2=\varepsilon_0+\frac{1}{2}\Gamma$.
\label{structure}}
\end{figure}

\begin{figure}
\centering \scalebox{0.35}{\includegraphics{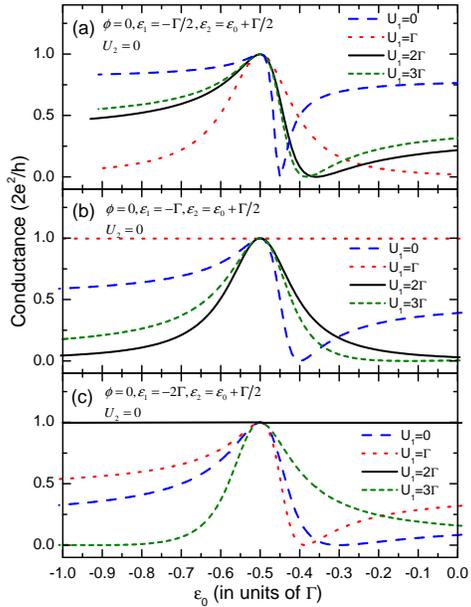}} \caption{
The conductance spectra influenced by the different-strength Coulomb
repulsions in QD-1, with the level of QD-1 being respectively fixed
at $-0.5\Gamma$, $-\Gamma$, and $-2\Gamma$. \label{highorder}}
\end{figure}

\begin{figure}
\centering \scalebox{0.35}{\includegraphics{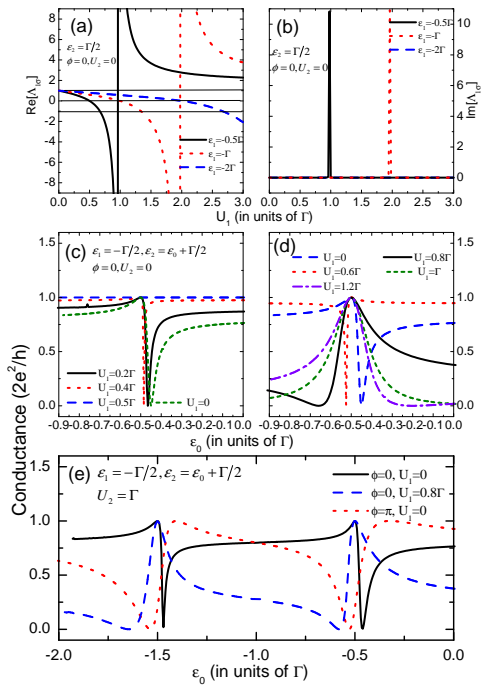}}
\caption{(a) and (b) The curves of $\rm{Re}\Lambda_{1\sigma}$ and
$\rm{Im}\Lambda_{1\sigma}$ as functions of $U_1$. (c) and (d) The
$U_1$-modulated conductance spectra. (e) The linear conductance in
the presence of nonzero $U_2$. \label{fanoline}}
\end{figure}

\begin{figure}
\centering \scalebox{0.35}{\includegraphics{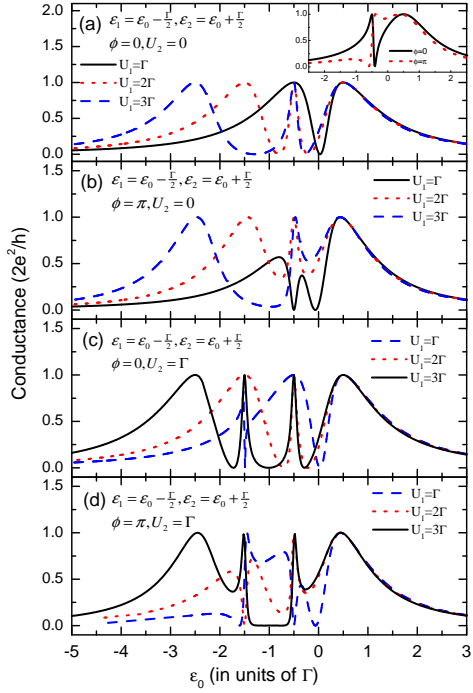}} \caption{
The Coulomb-modified conductance spectra for the cases of the
adjustable levels of QD-1 and QD-2. The relevant parameters are
taken as $\varepsilon_1=\varepsilon_0-\frac{1}{2}\Gamma$,
$\varepsilon_2=\varepsilon_0+\frac{1}{2}\Gamma$, and
$\Gamma_{11}=10\Gamma_{22}=\Gamma$. \label{fanoline1}}
\end{figure}

\begin{figure}
\centering \scalebox{0.35}{\includegraphics{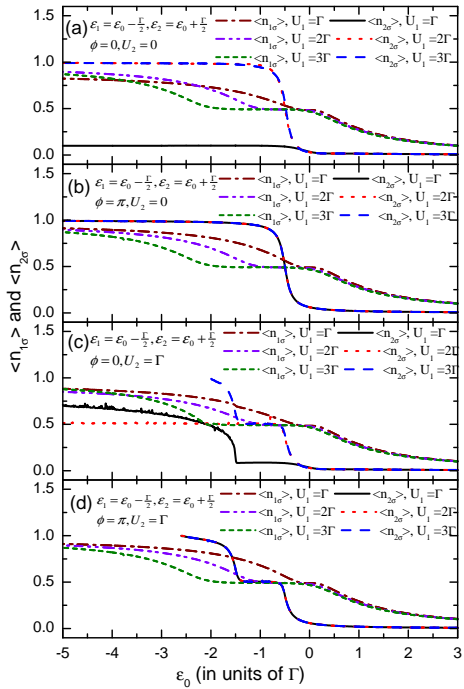}}
\caption{The spectra of Coulomb-modified average electron occupation
numbers in QD-1 and QD-2. The relevant parameters are the same as
those in Fig.\ref{fanoline1}. \label{orders}}
\end{figure}

\begin{figure}
\centering \scalebox{0.35}{\includegraphics{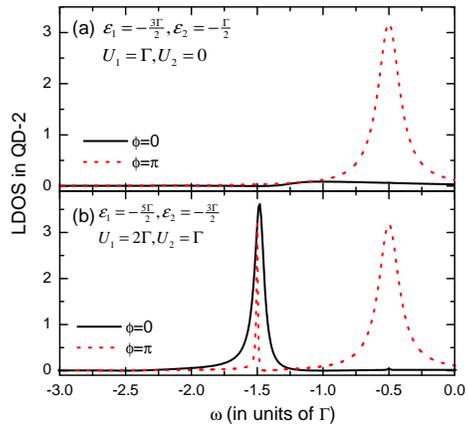}}
\caption{The spectra of LDOS in QD-2. The values of the relevant
parameters are assumed to be $\varepsilon_1=-\frac{3}{2}\Gamma$,
$\varepsilon_2=-\frac{1}{2}\Gamma$, $U_1=\Gamma$, and $U_2=0$ in
(a); $\varepsilon_1=-\frac{5}{2}\Gamma$,
$\varepsilon_2=-\frac{3}{2}\Gamma$, $U_1=2\Gamma$, and $U_2=\Gamma$
in (b). \label{compare}}
\end{figure}

\end{document}